\documentclass[aps,pre,twocolumn,superscriptaddress]{revtex4}
\usepackage{amsmath,amssymb}
\usepackage{graphicx}
\usepackage{subfigure}
\usepackage{color}

\newcommand{\sref}[1]{section~\ref{#1}}

\newcommand{\eref}[1]{(\ref{#1})}
\newcommand{\Eref}[1]{Equation~(\ref{#1})}

\newcommand{\fref}[1]{figure~\ref{#1}}
\newcommand{\Fref}[1]{Figure~\ref{#1}}

\newcommand{\erf}{\ensuremath{\mathop{\rm erf}}}
\newcommand{\erfc}{\ensuremath{\mathop{\rm erfc}}}

\newcommand{\vect}[1]{\mathbf{#1}}

\newcommand{\laplacian}{\vec{\nabla}^2}
\newcommand{\del}{\vec{\nabla}}

\newcommand{\singlet}[1]{\rho^{(1)}\left(\vect{#1}\right)}

\newcommand{\condsinglet}[2]{\rho\left(\vect{#1} \middle\vert \vect{#2}\right)}

\newcommand{\vdiff}[2]{\left|\vect{#1} - \vect{#2}\right|}
\newcommand{\ie}{\emph{i.e.}}
\newcommand{\eg}{\emph{e.g.}}

\newcommand{\phiR}{\ensuremath{\phi_{\rm R}}}

\newcommand{\V}{\ensuremath{\mathcal{V}}}
\newcommand{\Vr}{\ensuremath{\mathcal{V}_{\rm R}}}
\newcommand{\Vs}{\ensuremath{\mathcal{V}_0}}
\newcommand{\Vl}{\ensuremath{\mathcal{V}_1}}
\newcommand{\Vrl}{\ensuremath{\mathcal{V}_{\rm R1}}}

\newcommand{\lmf}{LMF}

\newcommand{\lj}{LJ}
\newcommand{\sig}{\ensuremath{\sigma}}
\newcommand{\sigmin}{\ensuremath{\sigma_{\rm min}}}
\newcommand{\vs}{\ensuremath{v_0(r)}}
\newcommand{\vl}{\ensuremath{v_1(r)}}
\newcommand{\us}{\ensuremath{u_0(r)}}
\newcommand{\ul}{\ensuremath{u_1(r)}}

\newcommand{\conddens}{\ensuremath{\condsinglet{\vect{r}^\prime}{\vect{r}}}}
\newcommand{\singdens}{\ensuremath{\singlet{\vect{r}}}}

\newcommand{\rhoqs}{\ensuremath{\rho^{q_\sigma}}}
\newcommand{\rhoq}{\ensuremath{\rho^q}}
\newcommand{\rhoG}{\ensuremath{\rho_G}}
\newcommand{\conv}{\ensuremath{\ast}}
\newcommand{\ybg}{YBG}

\newcommand{\kT}{\ensuremath{k_{\rm B}T}}

\begin{document}

\title{Local molecular field theory for the treatment of electrostatics}

\author{Jocelyn M. Rodgers}
\affiliation{Institute for Physical Science and Technology, University of
  Maryland, College Park, Maryland 20742}
\affiliation{Chemical Physics Program, University of Maryland, College Park, Maryland 20742}

\author{John D. Weeks}
\email{jdw@ipst.umd.edu}
\affiliation{Institute for Physical Science and Technology, University of
  Maryland, College Park, Maryland 20742}
\affiliation{Department of Chemistry and Biochemistry, University of Maryland,
  College Park, Maryland 20742}

\date{\today}

\begin{abstract}
  We examine in detail the theoretical underpinnings of previous
  successful applications of local molecular field (\lmf) theory to
  charged systems.  \lmf\ theory generally accounts for the averaged
  effects of long-ranged components of the intermolecular interactions
  by using an effective or restructured external field.
  The derivation starts from the exact Yvon-Born Green hierarchy and
  shows that the approximation can be very accurate when the
  interactions averaged over are slowly varying at characteristic
  nearest-neighbor distances.  Application of LMF theory to Coulomb
  interactions alone allows for great simplifications of the governing
  equations. LMF theory then reduces to a single equation for a
  restructured electrostatic potential that satisfies Poisson's
  equation defined with a smoothed charge density.  Because of this
  charge smoothing by a Gaussian of width \sig, this equation may be
  solved more simply than the detailed simulation geometry might
  suggest.  Proper choice of the smoothing length \sig\ plays a major
  role in ensuring the accuracy of this approximation. We examine the
  results of a basic confinement of water between corrugated wall and
  justify the simple \lmf\ equation used in a previous publication.
  We further generalize these results to confinements that include
  fixed charges in order to demonstrate the broader impact of charge
  smoothing by \sig.  The slowly-varying part of the restructured
  electrostatic potential will be more symmetric than the local
  details of confinements.
\end{abstract}

\maketitle

\section{Introduction \label{sxn:intro}}

The treatment of long-ranged Coulomb interactions remains a
substantial challenge in nearly all classical molecular simulations.
The basic problem is that interactions from even very distant charges
remain important, as indicated by the divergence of the integral of 1/r from any finite truncation distance to infinity. 
This causes problems in computer simulations of systems with charges, where contributions from distant periodic images of the
simulation cell must be taken into account, typically by Ewald sum techniques.
This also hampers the development of simple, intuitive local pictures and
analytic theory when charges are present.

Here we discuss basic features of a new approach for charged systems,
local molecular field (\lmf) theory.  \lmf\ theory was originally
applied to Lennard-Jones (\lj) systems for both simulation and
analytical
work~\cite{WeeksVollmayrKatsov.1997.Intermolecular-forces-and-the-structure-of-uniform-and-nonuniform,WeeksKatsovVollmayr.1998.Roles-of-repulsive-and-attractive-forces-in-determining,Vollmayr-LKatsovWeeks.2001.Using-mean-field-theory-to-determine,KatsovWeeks.2001.Density-fluctuations-and-the-structure-of-a-nonuniform-hard,KatsovWeeks.2001.On-the-mean-field-treatment-of-attractive-interactions,
Weeks.2002.Connecting},
and the \lmf\ approach has been used recently to analyze many charged
systems, both uniform and nonuniform, with even greater
success~\cite{ChenKaurWeeks.2004.Connecting-systems-with-short-and-long,ChenWeeks.2006.Local-molecular-field-theory-for-effective,RodgersKaurChen.2006.Attraction-between-like-charged-walls:-Short-ranged,DenesyukWeeks.2008.A-new-approach-for-efficient-simulation-of-Coulomb-interactions,RodgersWeeks.2008.Interplay-of-local-hydrogen-bonding-and-long-ranged-dipolar}.

LMF theory can be generally characterized as a mapping that relates
structural and thermodynamic properties of a nonuniform system with
long-ranged and slowly-varying intermolecular interactions in a given
external potential energy field,
representing, \eg, fixed solutes or confining walls,
to those of a simpler ``mimic system'' with properly chosen
short-ranged intermolecular interactions in an effective
or restructured field. The strong short-ranged interactions in the mimic system
are chosen to give an accurate description of the forces between
typical nearest-neighbor molecules in the full system, \eg, the
repulsive molecular cores in a dense LJ fluid, local hydrogen bonds in
water, or ion pairs in a strongly coupled ionic system. Because of the
short range of the interactions in the mimic system, fast and
efficient simulations scaling linearly with system size are possible.
The restructured field is determined self-consistently and is the sum of
the bare external field in the original system and a
density-weighted mean-field average of the remaining long-ranged and
slowly-varying parts of the intermolecular interactions. The restructured
field corrects major errors that can arise, especially in nonuniform
systems, from applying solely a spherical truncation of the
intermolecular pair interactions, exemplified by shifted force
truncations for nonuniform Coulomb
systems~\cite{FellerPastorRojnuckari.1996.Effect-of-Electrostatic-Force-Truncation-on-Interfacial,Spohr.1997.Effect-of-Electrostatic-Boundary-Conditions-and-System}.
More advanced spherical truncations of $1/r$ exist, such as
site-site reaction field
techniques~\cite{HummerSoumpasisNeumann.1994.Computer-simulation-of-aqueous-Na-Cl-Electrolytes},
but the difficulties in nonuniform situations remain the same.

The \lmf\ approach offers a very general perspective and the averaging
can be carried out for any slowly-varying component of the
intermolecular interactions.  We show here that LMF theory takes an
especially powerful and simple form, with strong analogies to
classical electrostatics, when it is applied uniformly to the basic
Coulomb $1/r$ interaction alone
(and not, \eg, to LJ interactions as well).  We can then take
advantage of symmetries between charges of different signs and
magnitudes, interacting with the same spherically symmetric $1/r$
potential, and determine the effective field using only the total
charge density, and not separate number densities for each charged
species or interaction site as would be required for a general
mixture.  We find that the contribution to the
restructured field in LMF
theory from the slowly-varying parts of the Coulomb interactions can
be exactly determined from a restructured electrostatic potential that
satisfied Poisson's equation, but with a Gaussian-smoothed charge
distribution.  A proper choice of the smoothing length $\sigma$ in the
Gaussian plays a key role in the accuracy of LMF theory, since it also
determines the size of the local strong-coupling region within which
strong forces from nearest-neighbor sites must be accurately captured
by the short-ranged interactions in the mimic system.

Previous discussions of \lmf\ theory for Coulomb or \lj\ systems relied on general arguments
and did not exploit all the important simplifications that arise by
applying LMF theory only to $1/r$ with the same potential splitting
regardless of point charge details.  Here we seek to justify and document
\lmf\ theory as completely as possible, accounting for Coulomb symmetries.
We show how this leads to a simple, broadly applicable
equation for a restructured electrostatic potential, and then we
extend results from~\cite{RodgersWeeks.2008.Interplay-of-local-hydrogen-bonding-and-long-ranged-dipolar}
to show that smooth \lmf\ solutions can still arise from a
general class of molecularly-corrugated confinement potentials.

\section{Derivation of the \lmf\ Equation \label{sxn:lmf}}
The \lmf\ equation may be derived for mixtures and standard site-site
molecular models.  However the basic features of the derivation are identical
regardless of whether we consider a mixture, a site-site molecular
fluid, or a single-component system.  Therefore, we present the
derivation for a single-component system in order to highlight the
physical approximations in the derivation without the clutter of the
multiple indices necessary in the derivations for mixtures or
molecules.  Furthermore, symmetries available to us when treating
charge interactions will allow us to simplify the \lmf\ equation for
mixtures into a single electrostatic potential equation, as we will
show later in \sref{sxn:lmfelect}.

\subsection{Motivation \label{sxn:LMFmotivation}}
We assume that the intermolecular interactions $w(r)$ in the full
system of interest are slowly varying at large separations.  The first
step in LMF theory is to divide $w(r)$ into properly chosen short- and
long-ranged parts:
\begin{equation}
  w(r) = \us + \ul,
  \label{eqn:potsplit}
\end{equation}
such that \ul\ contains all the the slowly-varying long-ranged
interactions in $w(r)$ but remains slowly varying at characteristic
nearest-neighbor distances where there are strong forces between
neighboring molecules.  By construction, the short-ranged component
\us\ then captures those strong short-ranged forces but vanishes
quickly at larger separations.

\lmf\ theory then defines a mapping from the full system with pair
potential $w(r)$ and an external single-particle potential energy
function $\phi(\vect{r})$ to a \emph{mimic} system defined by the
short-ranged pair interactions \us\ and an effective or restructured
external potential energy function $\phiR(\vect{r})$:
\begin{equation}
  \begin{array}{ccc}
    \mbox{Full} & & \mbox{Mimic}\,\, \\
    \left\{
      \begin{array}{c}
        w(r) \\
        \phi(\vect{r})
      \end{array} 
    \right\} &
  \stackrel{\mathrm{LMF}}{\rightarrow} &
  \left\{
    \begin{array}{c}
      \us \\
      \phiR(\vect{r})
    \end{array} 
  \right\}. 
  \end{array}
\label{eqn:LMFmap}
\end{equation}

We call the final combination a \emph{mimic} system when \us\ and
\phiR\ are properly chosen because we seek a system that can
capture many relevant structural and thermodynamic properties of the
full system.  In addition, we seek this mapping to be closed, in the
sense that the mapping should not require any information about the
full system other than the original $w(r)$ and $\phi(\vect{r})$.  It
seems plausible that some effective \phiR\ could account for averaged
effects of the neglected long-ranged interactions \ul\ in the mimic
system, but we seek a more careful development of this intuition.

\subsection{Exact Starting Point for the \lmf\ Derivation \label{sxn:ExactYBG}}
The derivation of the \lmf\ equation starts with an exact statistical
mechanical equation involving structure and forces, the
Yvon-Born-Green (\ybg) hierarchy of
equations~\cite{McQuarrie.2000.Statistical-Mechanics,HansenMcDonald.2006.Theory-of-Simple-Liquids}.
In one form, the first equation in the hierarchy reads
\begin{equation}
  \kT \del \ln \singlet{\vect{r}} = - \del \phi(\vect{r})
  - \int d\vect{r}^\prime \condsinglet{\vect{r}^\prime}{\vect{r}} \del
  w \left( \vdiff{r}{r^\prime} \right).
  \label{eqn:YBGeqn}
\end{equation}
This equation relates the gradient of the (singlet) density
profile of the system, \singdens, to the force from the external
single particle potential, $\phi(\vect{r})$, and the averaged force between particles,
weighted by the conditional singlet density, \conddens.  This function \conddens\ is
defined as the density at position $\vect{r}^\prime$ given that a
particle is at $\vect{r}$, and may be expressed mathematically as the
ratio of the two-particle density at $\vect{r}$ and $\vect{r^\prime}$
and the single-particle density at $\vect{r}$:
\begin{equation}
\conddens \equiv \frac{\rho^{(2)}(\vect{r},\vect{r}^\prime)}{\rho^{(1)}(\vect{r})}.
\label{eqn:conden}
\end{equation}
Thus, the \ybg\ hierarchy expresses the rapidly-varying singlet
density in terms of the more complex and even more rapidly-varying
two-particle density.  There are two related difficulties in trying to
extract practically useful information about the singlet density from
\eref{eqn:YBGeqn}.  First, the density profile is related to more
complicated rapidly-varying functions.  Second, there is no obvious
manner to create a self-contained loop to determine the singlet
density profile, again due to its relation to the higher-order density
profile.

Typical superposition closures of the \ybg\ hierarchy attempt to resolve both
difficulties by approximating $\condsinglet{\vect{r}^\prime}{\vect{r}}
\simeq \singlet{\vect{r}^\prime}$, leading to an equation where the
gradient over $\vect{r}$ can be pulled out of~\eref{eqn:YBGeqn} and
$\singlet{\vect{r}}$ is defined in terms of \emph{itself}.  However,
even for a bulk fluid, this is a poor approximation for any moderately dense
liquid~\cite{KrumhanslWang.1972.Superposition-assumption.-I.-Low-density-fluid-argon,WangKrumhansl.1972.Superposition-assumption.-2.-High-density-fluid-argon}.
The superposition approximation neglects any excluded volume or specific-binding
interactions that alter the small $\vdiff{r}{r^\prime}$ nature of
$\condsinglet{\vect{r}^\prime}{\vect{r}}$ relative to $\singlet{\vect{r}^\prime}$.
Replacing the conditional singlet
density with the singlet density simplifies the equation but also
effectively includes with large weight too many configurations where two particles are
improbably close to each other, leading to great inaccuracies.

We assume that there exists a good choice of \us\ that captures
those short-ranged excluded-volume and specific-binding contributions.  We will
describe such a
choice~\cite{ChenKaurWeeks.2004.Connecting-systems-with-short-and-long}
for splitting $1/r$ using the smoothing length \sig\ in
\sref{sxn:lmfelect}.  Given \us, we write the \ybg\ equations for both
the full sytem and the \lmf-mapped system, denoted by the subscript R, from \eref{eqn:LMFmap} as:
\begin{align}
&\kT \del \ln \singlet{\vect{r};[\phi]} = - \del \phi(\vect{r}) \nonumber \\
&\qquad \qquad - \int d\vect{r}^\prime \condsinglet{\vect{r}^\prime}{\vect{r};[\phi]} \del
  w \left( \vdiff{r}{r^\prime} \right), \nonumber\\
&\kT \del \ln \rho^{(1)}_{\rm R}\left(\vect{r};[\phiR]\right) = - \del \phiR(\vect{r}) \nonumber \\
&\qquad \qquad - \int d\vect{r}^\prime \rho_{\rm R}\left(\vect{r}^\prime\middle\vert\vect{r};[\phi_{\rm R}]\right) \del
  u_0 \left( \vdiff{r}{r^\prime} \right).
  \label{eqn:YBGboth}
\end{align}
The density profiles and conditional singlet densities are written as
functionals of $\phi(\vect{r})$ and $\phiR(\vect{r})$ to emphasize the
restructuring of the external potential energy in the LMF mapping.

As written with an arbitrary choice of \us,
\eref{eqn:YBGboth} simply displays formally exact but not
particularly useful equations for two arbitrary and unconnected
systems with different intermolecular interactions in different
external fields.  As in~\cite{WeeksKatsovVollmayr.1998.Roles-of-repulsive-and-attractive-forces-in-determining},
we now connect these systems and achieve substantial
simplifications and cancellations by requiring that \phiR\ be chosen
so that
\begin{equation}
  \rho^{(1)}(\vect{r};[\phi]) = \rho_{\rm R}^{(1)}(\vect{r};[\phiR]).
  \label{eqn:LMFSingletReq}
\end{equation}
It seems very plausible that such a \phiR\ exists, especially when
\us\ captures the strong short-ranged intermolecular forces.  We then take the
difference between the equations in \eref{eqn:YBGboth}.  The terms
involving the singlet densities on the left-hand sides cancel and the
result can be exactly written in a form useful for further analysis as
\begin{subequations}
\begin{align}
  \label{term:LMFMeanForce}
&  -\del \phiR(\vect{r}) = -\del \phi(\vect{r}) 
  - \int d\vect{r}^\prime \rho_{\rm R}(\vect{r}^\prime;[\phiR]) \del
  u_1\left(\vdiff{r}{r^\prime}\right) \\
  \label{term:MimicFullDiff}
&  \qquad - \int d\vect{r}^\prime  \left[ \rho(\vect{r}^\prime|\vect{r};[\phi])
    - \rho_{\rm R}(\vect{r}^\prime|\vect{r};[\phiR]) \right] \del
  u_0\left(\vdiff{r}{r^\prime}\right) \\
  \label{term:CondSingletDiff}
&  \qquad - \int d\vect{r}^\prime \left[\rho(\vect{r}^\prime|\vect{r};[\phi])
    -\rho(\vect{r}^\prime;[\phi]) \right] \del  
  u_1\left(\vdiff{r}{r^\prime}\right). 
\end{align}
\end{subequations}
Henceforth, for simplicity, we denote
$\singlet{\vect{r}}$ as $\rho(\vect{r})$.

Equation~(7) formally holds for any choice of \us\ and \ul, and because
conditional densities still appear, it generally has most of the problems
of the standard hierarchy equations noted after
\eref{eqn:conden}. Thus it may seem we have made little
progress. But this is not the case for proper choice of \us\ and \ul\
with the properties sketched above. This will allow considerable
simplification of~(7) and defines a mimic system that
\begin{itemize}
\item has a well-chosen \us\ that \emph{mimics} higher order
  correlations in the full system as well as the singlet density
  profile,
\item can be treated via statistical
  mechanics without any \emph{a priori} knowledge of the full system
  aside from $\phi(\vect{r})$ and $w(r)$, and
\item has an associated $\phiR(\vect{r})$ that depends only on
  singlet density profiles of the mimic system and not on the more
  complex conditional singlet densities.
\end{itemize}

\subsection{Approximations to Yield \lmf\ Equation \label{sxn:LMFapprox}}
Physically-motivated approximations to the exact~(7)
will result in a final \lmf\ equation below in
\eref{eqn:LMFGeneral} that has a simple mean-field form.  The
following discussion elucidates that the LMF equation is not a blind
assertion of mean-field behavior but rather a controlled and accurate
approximation, provided that we choose our mimic system carefully.

Recalling from \eref{eqn:potsplit} that $w(r) = u_0(r) + u_1(r)$, the
LMF derivation focuses on making two connected and physically-reasonable
approximations based on choosing a short-ranged \us\ that will induce
the correct nearest-neighbor structure, where the conditional singlet
and singlet densities differ the most, and a corresponding \ul\ 
that is slowly-varying on that length scale.  As we will see in
\sref{sxn:lmfpotslit}, the form of the Coulomb potential grants us
the freedom to fulfill both conditions by choosing a single smoothing
length \sig.  When $u_0(r)$ and $u_1(r)$ are chosen correctly, we
shall have a ``mimic system.''

By making approximations to~(7), the equality of
densities expressed in \eref{eqn:LMFSingletReq} will only hold
approximately, and we will find an equation based on the exact~(7)
that will also no longer be exact.  But all will be accurately
satisfied with proper choices of \us\ and \ul. The general
arguments below have been briefly sketched
before~\cite{WeeksKatsovVollmayr.1998.Roles-of-repulsive-and-attractive-forces-in-determining},
but the form of~(7) allows us to make a more precise statement of the necessary approximations.

We note that \eref{term:LMFMeanForce} depends only on the density
response of the mimic system, and no information about the density
response of the full system or any conditional density is needed.
Thus if terms~\eref{term:MimicFullDiff}
and~\eref{term:CondSingletDiff} could be neglected, then
\eref{term:LMFMeanForce} alone would define a closed equation for the
mimic system. The right side of \eref{term:LMFMeanForce} still
transforms as a gradient and should display all the associated
properties.  In that sense, neglecting terms~\eref{term:MimicFullDiff}
and~\eref{term:CondSingletDiff} together may well be more accurate
than taking either one of the following approximations individually.

With proper choice of \us\ and \ul, we may reasonably neglect the final
two terms in~(7) using the following arguments:
\begin{enumerate}
\item \label{item:SmallRApprox} Term~\eref{term:MimicFullDiff} probes
  the difference between the conditional singlet densities for the
  full and mimic systems via convolution with $\del \us$.  The
  integrand will be quickly forced to zero at larger
  $\vdiff{r}{r^\prime}$ by the vanishing gradient of the short-ranged
  $u_0\left(\vdiff{r}{r^\prime} \right)$.  Since both the full and
  mimic systems have the same strong short-ranged core forces with an
  appropriately chosen \us, and these core forces should mainly
  determine the short-ranged part of the conditional densities, it
  seems plausible that with proper choice of \us,
  term~\eref{term:MimicFullDiff} can be neglected. This is
  significantly better than any superposition approximation for the
  conditional densities and does not require us to numerically
  estimate either conditional density.
\item \label{item:LargeRApprox} Term~\eref{term:CondSingletDiff}
  probes the difference between the conditional singlet density and
  the singlet density of the \emph{full} system.  As explained
  previously with relation to standard closures of the YBG equation,
  assuming their equality can be highly problematic at short distances.  However, we are
  saved by the fact that this difference is paired with $\del
  u_1\left(\vdiff{r}{r^\prime}\right)$.  Since \us\ has been chosen to
  encompass core interactions, \ul\ will be simultaneously
  slowly-varying over those nearest-neighbor distances, so the
  associated force is essentially zero for exactly the range of
  distances where the conditional singlet density and singlet
  density differ significantly.  Thus we expect for many liquids that
  the integrand in term~\eref{term:CondSingletDiff} may also be
  accurately approximated as zero.
\end{enumerate}
These two approximations allow us to truncate the \ybg\ hierarchy.
No explicit knowledge of the conditional singlet density in either the full or
the mimic system will be required, but we expect the conditional
singlet density of the mimic system to closely resemble that of the
full system at short range.

Notably, the second approximation will fail when there are long-ranged pair
correlations arising from capillary waves at the liquid-vapor interface or the divergence of the
correlation length near the critical point.  In those instances, the
conditional singlet density will not be approximately the same as the singlet
density well beyond nearest-neighbor distances.  However, we do expect
this approximation to hold in strongly-coupled charged systems away from the critical point
where there exist pair correlations that decay exponentially over those
nearest-neighbor distances.

With these two coupled approximations, only \eref{term:LMFMeanForce}
remains, and we can take the gradient with respect to $\vect{r}$
outside of the integral.  Then, integrating over $\vect{r}$ yields the
general LMF equation
\begin{equation}
  \phiR(\vect{r}) = \phi(\vect{r}) + \int d\vect{r}^\prime 
  \rho_{\rm R}(\vect{r}^\prime;[\phiR])
  u_1\left(\left|\vect{r}-\vect{r}^\prime \right|\right) + C,
  \label{eqn:LMFGeneral}
\end{equation}
where the integration constant $C$ will be set by boundary conditions.
The form of this equation highlights the connection with mean-field
techniques.  The function \phiR\ may be viewed as an averaging of the
slowly-varying, long-ranged portions of the pair potential over the
single-particle density $\rho_{\rm R}(\vect{r})$.  The hierarchy has
been successfully truncated since \phiR\ is defined in terms of only
$\rho_{\rm R}(\vect{r})$.  Provided sufficient computer time to
simulate the short-ranged mimic system and therefore close the
self-consistent loop between $\phiR(\vect{r})$ and $\rho_{\rm
  R}(\vect{r})$, the \lmf\ equation is exactly soluble.

This equation is not a na\"ive mean-field ansatz for \phiR\ since
it arises from reasonable approximations to exact statistical
mechanical equations.  The crux of \lmf\ theory, however, is choosing
reasonable \us\ and \ul\ such that the necessary approximations are
accurate.  The next section discusses this choice for $1/r$.

\section{Electrostatics via LMF Theory \label{sxn:lmfelect}}
Here, we explain our choice of \us\ and \ul\ for charge interactions.
Then, we seek to highlight the symmetry and simplifications possible
when applying the \lmf\ equation to purely charge-charge interactions
by developing the \lmf\ equation for a single restructured electrostatic
potential, rather than for separate potential energy functions for
each atom type.

\subsection{Coulomb Potential Split \label{sxn:lmfpotslit}}
In CGS units, the pair
interaction between two charges, $q_\alpha$ and $q_\gamma$, is
\begin{equation}
  w(r) = \frac{q_\alpha q_\gamma}{\epsilon r}.
\end{equation}
The factor $\epsilon$ is simply the dielectric constant of a uniform
medium in which the charges are immersed.  The details of charge
magnitudes, signs, and choice of CGS or SI units will change
from system to system, but the underlying functional form $1/r$ will
remain constant.  Thus, 
following~\cite{DenesyukWeeks.2008.A-new-approach-for-efficient-simulation-of-Coulomb-interactions},
we introduce \vs\ and \vl\ to split this function as
\begin{equation}
  \frac{1}{r} \equiv \vs + \vl,
\end{equation}
into short-ranged and long-ranged components.  Both \vs\ and \vl\ will
yield either a potential energy or an electrostatic potential, with
appropriate prefactors.

Work of previous
researchers~\cite{ChenKaurWeeks.2004.Connecting-systems-with-short-and-long}
revealed that a beneficial choice of \vl\ is the functional form
associated with the electrostatic potential due to a unit Gaussian charge distribution of width
$\sigma$, defined as
\begin{equation}
  \rhoG(\vect{r}) = \frac{1}{\pi^{3/2}\sigma^3} \exp
  \left(-\frac{r^2}{\sigma^2}\right),
  \label{eqn:rhoG}
\end{equation}
leading to a \vl\ given by
\begin{equation}
\vl \equiv \int d\vect{r}^\prime \rhoG(\vect{r^\prime})\cdot \frac{1}{\vdiff{r}{r^{\prime}}}= \frac{\erf(r/\sigma)}{r}~.
  \label{eqn:v1ConvDef}
\end{equation}
This \vl\ is slowly-varying in $r$-space over the \emph{smoothing length} \sig.
As discussed in~\cite{ChenKaurWeeks.2004.Connecting-systems-with-short-and-long},
this choice of \vl\ also simultaneously isolates only the small wave-vector components of the Coulomb potential. 
The Fourier transform of \vl\ is
\begin{equation}
\hat{v}_1(k) = \frac{4\pi}{k^2} e^{-k^2\sig^2/4},
\end{equation}
and the Gaussian function attenuates any large-$k$ components that
would be rapidly-varying in $r$-space.  Thus the \vl\ is particularly
well-suited for mean-field averaging.

\begin{figure}[tb]
  \centering
  \includegraphics*[width=8cm]{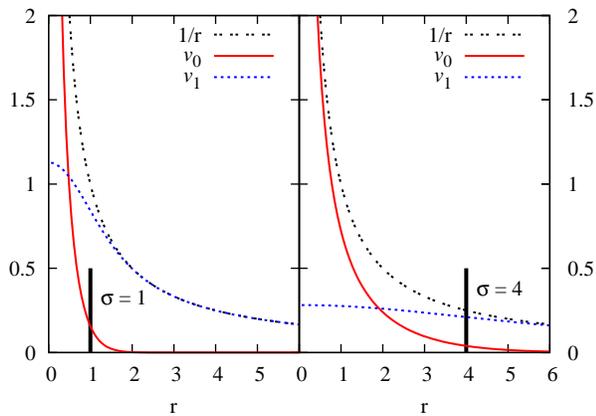}
  \caption{Demonstration of $1/r$ potential split for two different
    values of the smoothing length \sig.  When \sig\ is increased from
    1 to 4, more of the $1/r$ core interaction is included in \vs, and
    \vl\ becomes correspondingly more slowly varying.}
  \label{fig:PotSplit}
\end{figure}

The \vs\ corresponding to our choice of \vl\ will then be given as
\begin{equation}
  \vs\ = \frac{\erfc(r/\sigma)}{r}.
\end{equation}
Therefore, $v_{0}(r) \equiv 1/r - v_{1}(r)$ is proportional to the
electrostatic potential due to a point charge screened by a surrounding,
neutralizing Gaussian charge distribution whose width $\sigma$ also
sets the scale for the smooth truncation of \vs. This split is shown
in~\fref{fig:PotSplit} along with a demonstration of the impact of
increasing the smoothing length $\sigma$ on \vs\ and \vl.  The form of
\vl\ is slowly varying over \sig, and \vs\ approximately represents a
Coulomb core potential of range \sig.  Tuning \sig\ to larger values
increases the range of \vs\ and simultaneously smooths the variation
of \vl\ over characteristic nearest-neighbor interactions.

Since $1/r$ has no inherent length scale, the split between short- and
long-ranged by the smoothing length \sig\ may be tailored to each
situation.  This length \sig\ should not be viewed as a fitting
parameter, but rather a \emph{consistency} parameter.  We expect that
choosing \sig\ too small will result in poor and rapidly-varying
results from full LMF theory as $\sigma$ is changed since the
mean-field average will be over rapidly-varying interactions.  However
once \sig\ reaches a \sigmin\ that is related to characteristic
nearest-neighbor distances, the assumptions used in the previous
section to derive the \lmf\ equation from the \ybg\ hierarchy become
accurate, and essentially the same results should be found for larger
$\sigma$ values as well~\cite{oldfootnote}. Thus
we expect that self-consistently applied LMF theory with an
appropriately chosen \sig\ will yield a true
mimic system.  This requirement for reasonable core inclusion has been
explored in previous
papers~\cite{ChenKaurWeeks.2004.Connecting-systems-with-short-and-long,ChenWeeks.2006.Local-molecular-field-theory-for-effective,RodgersKaurChen.2006.Attraction-between-like-charged-walls:-Short-ranged}.

For small $r < \sigma$, the force due to \vs\ is quite similar to that
of a bare point charge.  By increasing $\sigma$, we increase the
effective range of these essentially unscreened Coulomb interactions.
Thus, \vs\ can be thought of as a Gaussian-truncated (GT) or ``short''
Coulomb interaction. We refer to a model with generically-truncated
Coulomb interactions as a short model and with \vs\ specifically used
as a GT model.  We showed
in~\cite{RodgersWeeks.2008.Interplay-of-local-hydrogen-bonding-and-long-ranged-dipolar}
that short GT water, where all the point charges in SPC/E water are replaced
by \vs\ with no account taken of the effective field, can give a
surprisingly accurate description of all site-site correlations for
bulk water. But this same approach can fail spectacularly for
electrostatic correlations in nonuniform systems unless corrected by
the inclusion of the long-ranged forces via the \lmf-restructured
potential~\cite{RodgersWeeks.2008.Interplay-of-local-hydrogen-bonding-and-long-ranged-dipolar}.

The functional form of \vs\ suggests both Ewald
sums~\cite{Ewald.1921.Evaluation-of-optical-and-electrostatic-lattice-potentials}
and Wolf
sums~\cite{WolfKeblinskiPhillpot.1999.Exact-method-for-the-simulation-of-Coulombic-systems}.
Ewald sums, however, seek to explicitly evaluate the forces from \vl\ at
each timestep using an exact sum over the fluctuating periodic images of the
simulation cell, and thus have a very different philosophical
approach.  The analogy with Wolf sums is much more appropriate in that
each results in a spherical truncation \vs\ applied at each time step.
However, the historical development of the Wolf summation technique
relies on the observation that charges in ionic lattices are quickly
neutralized by slightly shifted ``mirror'' lattices of neutralizing
charges.  In fluids, this is altered to assuming that charges within a
sharp cutoff radius are exactly neutralized by a shell of charge at the
cutoff radius, and the use of $\erfc(r/\sig)$ to damp the forces
from $1/r$ is applied more as a numerical afterthought.  In the
context of \lmf\ theory, we view such damping as much more crucial to
the success of Wolf sums in reproducing the structure of uniform
fluids than the imposition of strict neutrality within a sharp cutoff
radius.

The GT \vs\ also bears great similarity to site-site reaction field
(RF)
approaches~\cite{HummerSoumpasisNeumann.1994.Computer-simulation-of-aqueous-Na-Cl-Electrolytes,Neumann.1983.Dipole-Moment-Fluctation-Formulas-in-Computer,HummerSoumpasisNeumann.1992.Pair-correlations-in-an-NaCl-SPC-water-model,HummerSoumpasis.1994.Computation-of-the-Water-Density-Distribution-at-the-Ice-WaterInterface},
when the fluid is assumed to be conducting.  The usual RF method
generates a truncated $v_{0,{\rm RF}}(r)$ from the electrostatic
potential of a unit point charge surrounded by a uniform neutralizing
spherical charge distribution.  We discuss in the Appendix more
details of the relation between the GT \vs, and the usual RF
$v_{0,{\rm RF}}(r)$ as well as a smoother charged-clouds RF truncation
$v_{0,{\rm
    CC}}(r)$~\cite{HummerSoumpasisNeumann.1994.Computer-simulation-of-aqueous-Na-Cl-Electrolytes}.
The choice of spheres as the neutralizing charge distribution in these
RF methods leads to a potential that is exactly zero beyond a certain
cutoff distance, but this induces a discontinuity in a higher order
derivative at the sharp cutoff.  The \lmf\ choice of Gaussians leads to an
infinitely smooth and soft truncation both in $r$-space and
$k$-space. Moreover a GT model logically separates the choice of the
smoothing length $\sigma$ from the possible use for
computational convenience of a sharp cutoff at a larger distance along
with various local smoothing schemes that could be employed near the
cutoff.

Despite the general similarities, the theoretical constructs of both reaction
field techniques and Wolf sums allow for no obvious correction in
nonuniform systems, such as the slab confinement to which we applied
\lmf\ theory in~\cite{RodgersWeeks.2008.Interplay-of-local-hydrogen-bonding-and-long-ranged-dipolar}.
As such, \lmf\ theory is an important advance in the use of spherical
truncations for $1/r$.

\subsection{\lmf\ Electrostatic Potential \label{sxn:lmfelectpot}}
Now we will derive a further simplified electrostatic form of the
\lmf\ equation.  We need only consider the \lmf\ mixture equation here
because, with appropriate further approximations, the \lmf\ equation
for standard molecular models collapses onto the mixture \lmf\ 
equation as will be shown in a paper dealing more broadly with
site-site molecular models~\cite{HuRodgersWeeks..}.  Using the \ybg\ 
equation for mixtures, exactly the same approximations as in
\sref{sxn:lmf} lead to the following \lmf\ equation for a species
$\alpha$ in a mixture of different species indexed by $\gamma$ and
including $\alpha$, as originally given
in~\cite{ChenWeeks.2006.Local-molecular-field-theory-for-effective}:
\begin{equation}
 \phi_{R,\alpha} = \phi_{\alpha}(\vect{r}) + 
  \sum_{\gamma} \int d\vect{r}^\prime \, \rho_{R,\gamma}(\vect{r}^\prime) \cdot u_{1,\alpha\gamma}(\vdiff{r}{r^\prime}).
  \label{eqn:LMFmixture}
\end{equation}
If the pair interactions being treated were of general form this
would be the furthest simplification possible, with an \lmf\ equation
to be solved for each different species.

The observation
in~\cite{ChenWeeks.2006.Local-molecular-field-theory-for-effective}
that a single \sig\ for all charged interactions proves most useful
leads straightforwardly to a much more compact form.  Using the \lmf\ 
approach only for charges, each $u_{1,\alpha\gamma}(r)$ may be written
as
\begin{equation}
  u_{1,\alpha\gamma}(r) = \frac{q_\alpha q_\gamma}{\epsilon}\frac{\erf\left(r/\sig\right)}{r} = \frac{q_\alpha q_\gamma}{\epsilon} v_1(r).
\end{equation}
Thus \eref{eqn:LMFmixture} may be exactly written as
\begin{equation}
  \phi_{{\rm R},\alpha}(\vect{r}) = \phi_{\alpha}(\vect{r}) +
  \frac{q_\alpha}{\epsilon} \int d\vect{r}^\prime \, \left(
    \sum_{\gamma} q_\gamma \rho_{{\rm R},\gamma}(\vect{r}^\prime) \right)
  \cdot v_{1}(\vdiff{r}{r^\prime}).
\end{equation}
Using the natural definition of charge density as
\begin{equation}
    \rhoq(\vect{r})  = \sum_{\gamma} q_\gamma \rho_{\gamma}(\vect{r}),
\end{equation}
we may reexpress the previous equation for $\phi_{{\rm R},\alpha}$ as
\begin{equation}
  \phi_{{\rm R},\alpha}(\vect{r}) = \phi_{\alpha}(\vect{r}) + \frac{q_\alpha}{\epsilon} \int
  d\vect{r}^\prime \, \rhoq_{\rm R}(\vect{r}^\prime) \cdot
  v_{1}(\vdiff{r}{r^\prime}).
\end{equation}
By noting as done in \cite{ChenWeeks.2006.Local-molecular-field-theory-for-effective}
that $\phi_{\alpha}$ may be due to both electrostatic and nonelectrostatic components, we separate $\phi_{\alpha}$ as
\begin{equation}
  \phi_{\alpha}(\vect{r}) = \phi_{{\rm ne},\alpha}(\vect{r}) + q_\alpha \mathcal{V}(\vect{r})
\end{equation}
where $\phi_{{\rm ne},\alpha}$ encompases nonelectrostatic confinements
such as the smooth \lj\ walls used in~\cite{RodgersWeeks.2008.Interplay-of-local-hydrogen-bonding-and-long-ranged-dipolar} and
\V\ represents the general external electrostatic potential.

This observation naturally leads to the following rewriting of each \lmf\ equation for a species $\alpha$ as
\begin{equation}
  \phi_{{\rm R},\alpha}(\vect{r}) = \phi_{{\rm ne},\alpha}(\vect{r}) + q_\alpha \Vr(\vect{r}),
\end{equation}
with a single restructured electrostatic potential \Vr, defined as
\begin{equation}
  \Vr(\vect{r}) = \V(\vect{r}) + \frac{1}{\epsilon} \int d\vect{r}^\prime \rhoq_{\rm R}(\vect{r}^\prime) \cdot v_1(\vdiff{r}{r^\prime}).
\label{eqn:rescaledV}
\end{equation}
Thus all self-consistent requirements on the diverse set of
$\phi_{{\rm R},\alpha}$ can be written in terms of a single restructured
electrostatic potential \Vr\ defined by one \lmf\ equation involving
the equilibrium mobile charge density \rhoq.

Given the convolution definition of \vl\ as the potential due to a
Gaussian charge density in \eref{eqn:v1ConvDef}, we may schematically represent the
integral term in \eref{eqn:rescaledV} as $\rho^q$\conv\rhoG\conv$1/r$
where \conv\ indicates a convolution.
Thus \eref{eqn:rescaledV} can be exactly rewritten  as
\begin{equation}
  \Vr(\vect{r}) = \V(\vect{r}) + \frac{1}{\epsilon} \int d\vect{r}^\prime \, \rhoqs_{\rm R}(\vect{r}^{\prime}) \cdot
  \frac{1}{\vdiff{r}{r^{\prime}}},
\label{eqn:smoothedrhomobile}
\end{equation}
where \rhoqs\ represents the smoothing of the charge density
by \rhoG\ in \eref{eqn:rhoG} via the convolution
\begin{equation}
  \rhoqs(\vect{r}) \equiv \int d\vect{r}^\prime \, \rhoq(\vect{r}^\prime) \rhoG(\vdiff{r}{r^\prime}).
\label{eqn:conv}
\end{equation}

In almost all cases of interest we can picture $\mathcal{V}$ as
arising from a \emph{fixed}, external charge distribution $\rhoq_{\rm
  ext}$, and we should apply the same separation of the Coulomb
potential to the fixed as well as mobile charges. This will allow us
to express \eref{eqn:smoothedrhomobile} in a suggestive and compact
form using the sum of the fixed charge density and the equilibrated
mobile charge density profile, $\rhoq_{\rm tot}(\vect{r}) =
\rhoq(\vect{r}) + \rhoq_{\rm ext}(\vect{r})$. Thus we write
\begin{equation}
  \V(\vect{r}) = \Vs(\vect{r}) + \Vl(\vect{r}).
\end{equation}
Just as \vl\ is the electrostatic potential due to a point charge
convolved with a Gaussian of width \sig, so \Vl\ is the electrostatic
potential due to the convolution of the \emph{fixed} external charge
density $\rhoq_{\rm ext}$ with that same Gaussian as defined in
\eref{eqn:rhoG} and used in \eref{eqn:v1ConvDef} and \eref{eqn:conv}.
Given the definition of \Vs\ and \Vl, we write analogously
\begin{equation}
  \Vr(\vect{r}) = \Vs(\vect{r}) + \Vrl(\vect{r}),
\end{equation}
thus defining $\Vrl(\vect{r})$, the slowly-varying component of the restructured electrostatic
potential. 

Using \eref{eqn:smoothedrhomobile}, this slowly-varying component
is exactly given by
\begin{equation}
  \Vrl(\vect{r})  = \frac{1}{\epsilon} \int
  d\vect{r}^{\prime} \, \rhoqs_{\rm R,tot}(\vect{r}^{\prime}) \cdot
  \frac{1}{\vdiff{r}{r^{\prime}}} \label{eqn:LMFsmoothq}.
\end{equation}
We see that \Vrl\ is the electrostatic potential arising from the
Gaussian smoothing of the \emph{total} charge density, $\rhoqs_{\rm
  tot}(\vect{r})$.

\Eref{eqn:LMFsmoothq} appears as the standard definition of the
electrostatic potential due to charge density from first-year
textbooks like \cite{Purcell.1985.Electricity-and-Magnetism}.  Thus
the \lmf\ equation for the slowly-varying component of the restructured
electrostatic potential may be viewed as simply doing classical
electrostatics on an \emph{equilibrium}, \emph{smoothed} total charge
density and defining a single external restructured electrostatic
potential.  In fact we may in general represent the slowly-varying
\lmf\ equation \eref{eqn:LMFsmoothq} as a modified Poisson's equation,
\begin{equation}
  \laplacian \Vrl(\vect{r})= - \frac{4\pi}{\epsilon} \rhoqs_{\rm R,tot}(\vect{r}),
\end{equation}
based on the smoothed total charge density.  This point was first made for
the uniformly-charged-wall model
system~\cite{Chen.2004.General-Theory-of-Nonuniform-Fluids:-From}, and
it is immediately generalizable to the \lmf\ equation dealt with
here.  The existence of this Poisson-like equation dictating a modified
  electrostatics emphasizes that, in order for \lmf\ theory to be valid
  for electrostatics, \emph{all} charge-charge interactions, whether fixed, bound, or mobile, should be
  mapped via \lmf\ theory to short-ranged interactions in the
  restructured electrostatic potential \Vr.
In~\cite{RodgersWeeks.2008.Interplay-of-local-hydrogen-bonding-and-long-ranged-dipolar},
we emphasized the value of this modified-electrostatics perspective in
physically interpreting molecular simulations and in understanding why
and when spherical Gaussian truncations, using \vs\ alone, fail in
nonuniform environments.

\section{Further Simplifications Due to Charge Smoothing \label{sxn:chgsmooth}}
Understanding the \lmf\ equation as standard electrostatics for a
Gaussian-smoothed charge density also allows for simpler solutions of
the \lmf\ equation than might be initially expected.  As an example,
in~\cite{RodgersWeeks.2008.Interplay-of-local-hydrogen-bonding-and-long-ranged-dipolar},
we were able to use a one-dimensional \lmf\ equation to treat a system
that has a charge density profile with an explicitly three-dimensional
character.  In this section, we give the reasoning behind such
simplifications and argue that they should be expected generally in
many physically relevant cases.

\subsection{Simple Corrugated Surface \label{sxn:simplecorr}}
In~\cite{RodgersWeeks.2008.Interplay-of-local-hydrogen-bonding-and-long-ranged-dipolar},
we treated SPC/E water confined to a slab by two empirical Pt(111)
surfaces~\cite{RaghavanFosterMotakabbir.1991.Structure-and-Dynamics-of-Water-at-the-Pt111-Interface:}
that order the water molecules at the surface, attracting the oxygen
atoms to localized binding sites. %
Details of the simulation methods are available
  in~\cite{RodgersWeeks.2008.Interplay-of-local-hydrogen-bonding-and-long-ranged-dipolar}.
  In brief, we simulated 1054 SPC/E molecules confined between the two
  walls with 500 ps of equilibration and 1.5 ns of subsequent data
  collection.  Simulations using slab-corrected three-dimensional
  Ewald sums, LMF theory with $\sigma = 4.5$~\AA, and LMF theory with
  $\sigma=6.0$~\AA\ were carried out. During those runs, the timestep
  chosen was 1.0 fs.  However, for self-consistent solution of the LMF
  equation, we first simulated sets of 10 shorter runs with 25 ps of
  equilibration and 50 ps of accumulation with a timestep of 2.5 fs.
  These 10 shorter runs began with distinct initial conditions; we
  observed that averaging the \rhoq\ resulting from such a set of
  simulations better isolated the system's equilibrium response to
  changes in \Vr\ from the natural fluctuations in the charge density.
  For the larger $\sigma$, convergence was achieved within 3
  iterations, and for the smaller $\sigma$, 10 iterations were
  required for self-consistency.  This approach certainly does not represent a
  numerically optimized solution of the LMF equation and is used here only
  to demonstrate the accuracy possible in principle when the full LMF theory is
  used. We will address the need for a more comprehensive
  and numerically efficient approach to the solution of the
  \lmf\ equation and general timing issues further in \sref{sxn:conc}.
  
  \begin{figure}[tb]
  \centering
  \subfigure[]{
    \includegraphics*[width=3.3cm]{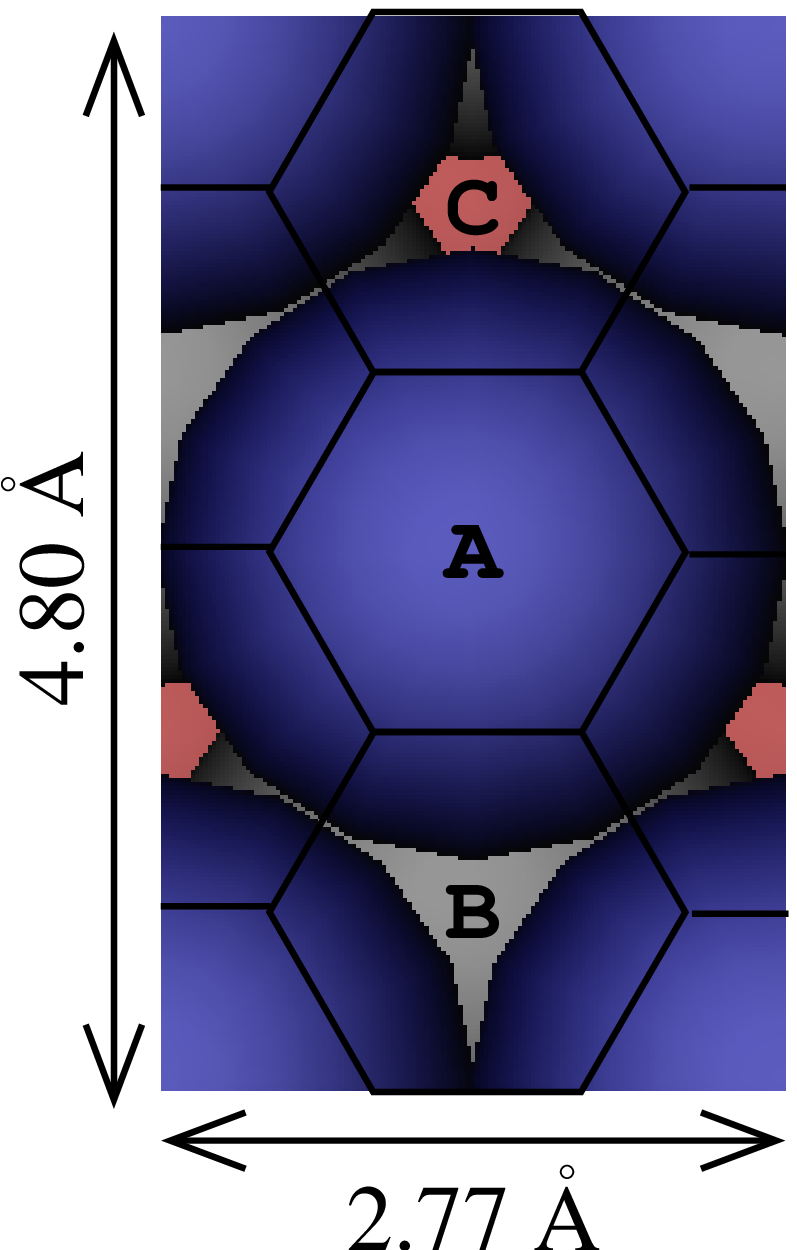}
    \label{fig:Ptunit}
  } 
  \subfigure[]{ 
    \includegraphics*[width=4.5cm]{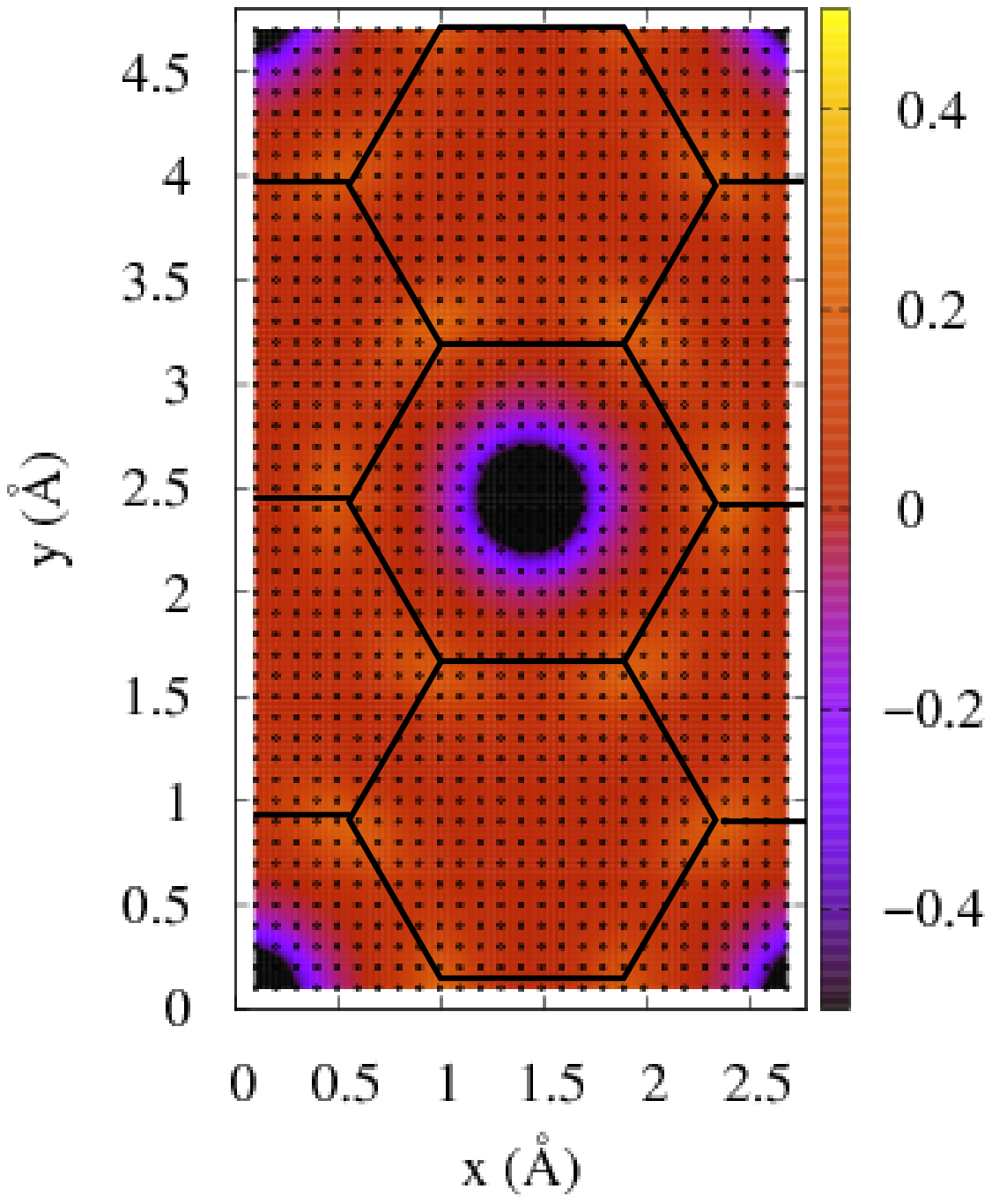}
    \label{fig:PtChgPlot}
  }
  \caption{A (111) surface and the proximal charge density. \subref{fig:Ptunit}~Sketch of the rectangular
    unit cell for a (111)~surface showing the topmost three layers of
    atoms, labelled A, B, and C.  \subref{fig:PtChgPlot}~Corresponding
    $\rhoq(x,y)$ in e$_0$/\AA$^3$ for the 1.5~\AA\ layer closest to
    the wall.  The density is projected onto the $xy$-plane, and the
    grayscale (color online) indicates the value of the charge at each
    point.  Black dots indicate collected data points.}
  \label{fig:Ptsurface}
\end{figure}

In contrast to the smooth hydrophobic surface also dealt with in that
paper, the empirical (111) surface was intended to represent detailed
and specific ordering of water molecules at an atomically-corrugated
surface.  A schematic of the rectangular unit cell of a (111) surface
is shown in \fref{fig:Ptunit}, demonstrating the spacing of the three
topmost layers of atoms.  The empirical Pt(111)
potential~\cite{RaghavanFosterMotakabbir.1991.Structure-and-Dynamics-of-Water-at-the-Pt111-Interface:}
that we use does not have explicit atoms or charges. Rather the
potential simply encompasses the presence of binding sites through
exponential functions and various cosine functions to represent the
periodicity of the surface.

As expected from the form of the potential, the charge density \rhoq\ 
depends on $x$ and $y$ in addition to $z$.  In \fref{fig:PtChgPlot},
the charge density $\rhoq(x,y)$ in the 1.5 \AA\ layer nearest the
(111)-wall is projected onto a unit cell.  As previously
demonstrated~\cite{RaghavanFosterMotakabbir.1991.Structure-and-Dynamics-of-Water-at-the-Pt111-Interface:},
the surface induces a charge density profile very nonuniform in the
$x$- and $y$-directions.  The distinct periodicity is due to the fact
that the spacing between binding sites (2.77~\AA) is nearly
commensurate with the first peak in the bulk water $g_{\rm OO}(r)$
($\sim$ 2.8~\AA).  However despite a charge density that was
explicitly a function of $\vect{r}$, we obtained accurate results for
system properties using
$\Vr(z)$~\cite{RodgersWeeks.2008.Interplay-of-local-hydrogen-bonding-and-long-ranged-dipolar}.
We posited that this was the case due to the charge smoothing
described in the previous section.

\begin{figure}
  \centering \includegraphics*[width=8cm]{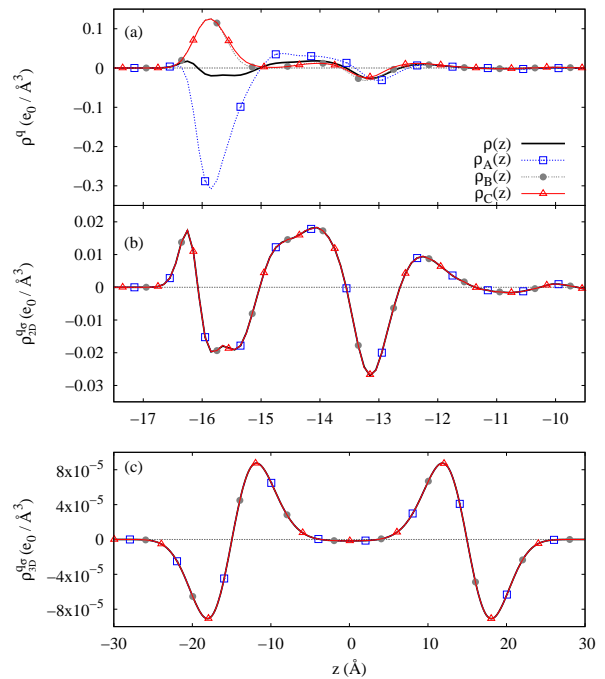}
  \caption{Effect of smoothing with a Gaussian of width $\sig=4.5$~\AA\ on $\rhoq(\vect{r})$
 between two empirical (111) walls.  (a)~The charge density $\rhoq(z)$ for the three hexagonal
prisms labeled A, B, and C in \fref{fig:Ptunit}.  (b)~Gaussian smoothing of these density profiles
in the $x$- and $y$-directions.  (c)~Full three-dimensional Gaussian smoothing of the $\rhoq(z)$ profiles.
The plots in (a) and (b) focus on data near the left wall, but the plot in (c) displays data for the full slab of water between the walls.}
  \label{fig:PtSmoothChg}
\end{figure}

Here we seek to demonstrate this more quantitatively by probing the
charge density profile away from the surface within hexagonal prisms
defined by three distinct hexagonal binding regions -- A, B, and C -- as
labeled in \fref{fig:Ptunit}.  \Fref{fig:PtSmoothChg}(a) shows the
$\rhoq(z)$ measured along each hexagonal prism and averaged laterally over
the hexagonal area.  These density profiles are quite distinct both from each
other and from the full laterally-smoothed $\rhoq(z)$.

The Gaussian smoothing of charge density inherent in the LMF treatment
allows us to distinguish the coherent and incoherent variations in
charge density seen in \fref{fig:PtSmoothChg}(a).  As shown in
\fref{fig:PtSmoothChg}(b), convoluting $\rho^{q}(\vect{r})$ with
Gaussians in the $x$- and $y$-directions and subsequently analyzing
sites A, B, and C leads to a density, $\rho^{q_\sigma}_{2D}$, that is
indistinguishable from the $\rho^{q}(z)$ calculated initially as only
a function of $z$.  This equivalence is due to that fact that the
corrugations in effect induce local random noise in the density
profile that is therefore averaged out by the Gaussians of width
greater than the corrugation.  In contrast, the presence of the
confining surface leads to a $\rho^{q_\sigma}$ density profile that
does not average out in the $z$-direction as seen in
\fref{fig:PtSmoothChg}{c}.  In essence, the \lmf\ smoothing approach
shows when variations in charge density profiles cancel, and when they
do not.

\subsection{Simplifications Still Hold For Surfaces with Fixed Charges \label{sxn:surfchg}}
However, a reasonable criticism of the previous surface is that it
was not atomically realistic, even on a classical level.  
More realistic metallic surfaces would require a treatment of image
charges, which we will not address here.  However, molecular surfaces
modeled with standard force fields would at least have charges at
atomic sites near the surface.  If the charges are free to move
throughout the simulation as would generally be the case for
biological membranes or liquid-liquid interfaces, absolutely no change
in tactic would be necessary.  The charges associated with those
molecules would also be included in the mobile \rhoq.

If, instead, the charges are held fixed, as is more likely for solid
surfaces in classical simulations of solid-liquid interfaces, then
$\phi_\alpha(\vect{r}) = \phi_{\alpha, {\rm ne}}(\vect{r}) + q_\alpha
\V(\vect{r})$.  In such cases, we may no longer approximation
$\Vr(\vect{r})$ as $\Vr(z)$, but a similar simplification for the
smooth \Vrl\ will hold, as might be expected.
Such a potential for silica surfaces~\cite{GiovambattRosskyDebenedett.2006.Effect-of-pressure-on-the-phase-behavior-and-structure} is currently being
employed in our group to study the silica-acetonitrile interface~\cite{HuUnpub}.

For surfaces with fixed charges, the imposed external electrostatic potential
$\mathcal{V}(\vect{r})$ would be
\begin{equation}
  \mathcal{V}(\vect{r}) = {\sum_{{\rm sites} \, i}}^\prime \frac{q_i}{\epsilon \vdiff{r}{r_i}},
\end{equation}
where the prime indicates that these surface charges will need to be
accounted for in the surface extending to infinity in the $x$- and
$y$-directions.
The pair interactions representing the excluded volume of these atomic
sites would be included in $\phi_{{\rm ne},\alpha}$.  The external electrostatic
potential \V\ may be split into short-ranged component \Vs\ and
long-ranged component \Vl\ as
\begin{eqnarray}
  \Vs(\vect{r}) = \sum_{{\rm sites} \, i} \frac{q_i}{\epsilon} v_0(\vdiff{r}{r_i}),\\
  \Vl(\vect{r}) = {\sum_{{\rm sites} \, i}}^\prime \frac{q_i}{\epsilon}
  v_1(\vdiff{r}{r_i}).
\end{eqnarray}
Now \Vs\ is a simple minimum image sum over short-ranged pair
interactions.  The long-ranged component \Vl\ still must include the
effect of surface sites extending out to infinity in the $x$- and
$y$-directions, but since \Vl\ may be interpreted as the electrostatic
potential due
to a Gaussian smoothing of the fixed charge sites, we expect that
\begin{equation}
  \Vl(\vect{r}) \simeq \Vl(z).
\end{equation}
The sole requirement would be that \sig\ should be larger than
relevant lateral spacing of sites on the surface, such that the charge
density will be sufficiently smoothed in the lateral directions.  In
such a case, \Vr\ is a function of $\vect{r}$, but all dependence on
$x$ and $y$ is contained in $\Vs(\vect{r})$, and we would expect
the self-consistent $\Vrl(z)$ to hold to a very good approximation.

As a quick demonstration of the validity of these ideas, we
construct a (111)-surface with lattice constant $a = 3.92$~\AA\ as for
the model (111)-surface discussed in \sref{sxn:simplecorr}.  However,
we arbitrarily assign charges of $+1$~e$_0$ for the first layer of
atoms, $-2$~e$_0$ for the second layer of atoms, and $+1$~e$_0$ for
the third layer of atoms.

\begin{figure}[tb]
  \centering
  \includegraphics*[width=8cm]{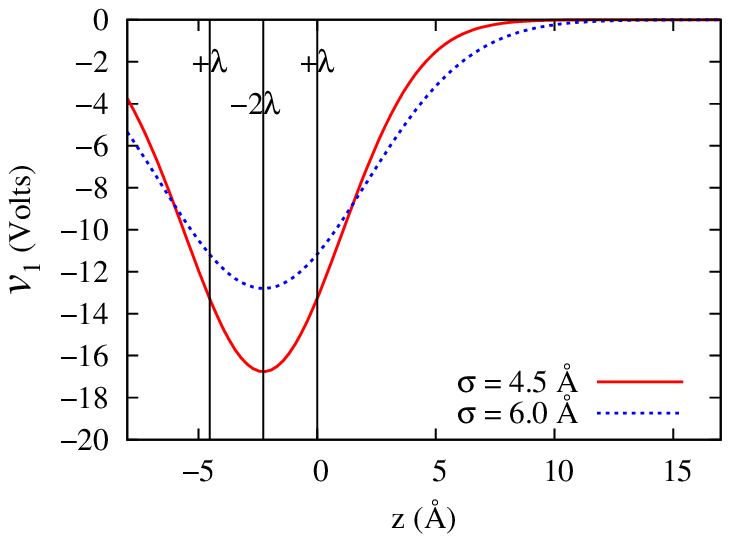}
  \caption{$\Vl(z)$ for charged (111)-surface for two different \sig.
This approximation to $\Vl(\vect{r})$ is nonnegligible within $2\sig$ of the surface.}
  \label{fig:Vl111Surface}
\end{figure}

The corresponding $\Vl(z)$ would be determined by smoothing out the
charges in each layer laterally, such that the charged (111)-surface
is now approximated by three uniformly charged planes.  With this
representation, we may exactly express $\Vl(z)$ as the sum of three 
analytical functions,
\begin{equation}
  \Vl(z) = \sum_{i \in {\rm layers}} \frac{2 \pi \lambda_i}{\epsilon} G(z,z_i),
\label{eqn:V1z}
\end{equation}
where $\lambda_i$ is the surface charge density of each layer, and $G$
is the one-dimensional Gaussian-smoothed Green's function
for a planar charge distribution \cite{ChenWeeks.2006.Local-molecular-field-theory-for-effective,RodgersKaurChen.2006.Attraction-between-like-charged-walls:-Short-ranged}
analogous to \vl\  in Eq.\ \eref{eqn:v1ConvDef} for a point charge.

$\Vl(z)$ is plotted in \fref{fig:Vl111Surface} and is an important
contributor to the forces near the surface.  However, since the
surface is net neutral, it decays essentially to zero within a distance of
approximately $2\sig$ from the surface.

We may calculate both $\V(\vect{r})$ and $\Vs(\vect{r})$ easily by
constructing various configurations of one $+1$ unit test charge
above this (111)-surface in the {\sc dlpoly}2.16 molecular dynamics
package.  We exactly calculate $\Vs(\vect{r})$ to within the
simulation energy precision, by carrying out single point energy
calculations for a $+1$~e$_0$ test charge above various $(x,y)$-sites
interacting with each particle in the surface via \vs.  This particle
is a test charge in the sense that it measures the \Vs\ generated by
the charges in the surface without disturbing the organization of
charges within the surface.  Thus the \lmf\ estimation of the total
$\V(\vect{r})$ is
\begin{equation}
  \V(\vect{r}) \simeq \V_{\rm LMF}(\vect{r}) \simeq \Vs(\vect{r}) + \Vl(z),
\end{equation}
with $\Vl(z)$ given analytically by \eref{eqn:V1z}.

In order to determine $\V(\vect{r})$ independently, we employ slab-corrected
three-dimensional Ewald sums as a way to estimate that
function~\cite{YehBerkowitz.1999.Ewald-summation-for-systems-with-slab}.
Since Ewald sums require net neutrality, we construct two surfaces
with test charges -- the surface we are interested in with a $+1$ test
charge above it and a second mirror surface with opposite charges with
a $-1$ test charge above it. The two (111)-surfaces have lateral size
30.49~\AA~$\times$~28.806~\AA\ with $L_z = 280$~\AA, and we use
$\alpha = 0.34$~\AA$^{-1}$ and $\vect{k}_{\rm max} = (13,12,120)$ for
the corrected three-dimensional Ewald sums.
We separate the two surfaces within the simulation cell as much as
possible in order to decouple each surface and test-charge pair from
the other pair.

The energies yielded from these single-point-energy calculations
$U_{\rm Ewald}(\vect{r})$ allow us to estimate $\V(\vect{r})$,
provided that we subtract the contribution from test charge-test
charge interactions $U_{+-}(z)$ and the constant contribution from
intra-wall particle interactions ($U_{\rm wall}$) as follows,
\begin{equation}
  \V(\vect{r}) \simeq \V_{\rm Ewald}(\vect{r}) =  \frac{1}{2} \left\{ U_{\rm Ewald}(\vect{r}) - U_{+-}(z) - U_{\rm walls} \right\}.
\end{equation}
This is approximate since the Ewald sums will propagate the test
charges laterally into all the periodic image cells, but given that
the lateral area spans $11 \times 6$ repeats of the minimum
rectangular (111)-surface unit cell of 2.77~\AA\ $\times$ 4.80~\AA,
interactions between each surface and the lateral periodic images of
the nearest test charge should be minimal.

\begin{figure}[tb]
  \centering
  \includegraphics[width=8cm]{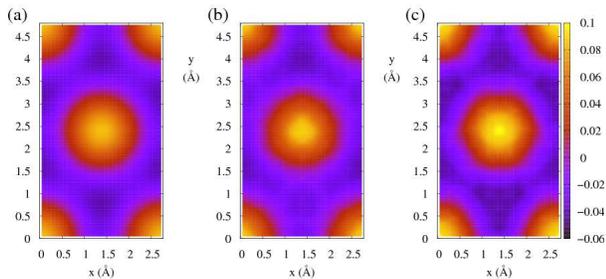}
  \caption{Electrostatic potential $\V(\vect{r})$ at a distance of 2.26~\AA\ above the charged
surface as calculated by (a)~corrected three-dimensional Ewald sums and by \lmf\ estimation
with (b)~$\sig = 4.5$~\AA\ and (c)~$\sig = 6.0$~\AA.  \lmf\ estimation entails
approximating $\V(\vect{r})$ as $\Vs(\vect{r}) + \Vl(z)$.}
  \label{fig:VEwaldSurface}
\end{figure}

Shown in~\fref{fig:VEwaldSurface} is $\V(x,y,z)$ for $z$ fixed at
$2.26$~\AA\ as treated by (a)~Ewald sums, (b) the \lmf\ approximation
with $\sig=4.5$~\AA, and (c)~the \lmf\ approximation with
$\sig=6.0$~\AA.  The similarity of the electrostatic potential surface for each
\sig\ to the full Ewald calculation demonstrates that 4.5~\AA\ is
greater than \sigmin\ for the fixed surface at least.  This stands to
reason since, within a plane, particles are within 2.77~\AA\ of each
other and the planes of particles are separated by about 2.26~\AA.
Therefore $\sig=4.5$~\AA\ is greater than relevant particle spacings.
The agreement between $\V_{\rm LMF}(\vect{r})$ and $\V_{\rm
  Ewald}(\vect{r})$ is expected but numerically nontrivial since
$\left| \Vl \right| \gg \left| \V \right|$ at this distance above the
surface.  Therefore, achieving this degree of accuracy is a result of
the accurate addition of two large-magnitude terms, $\Vs(\vect{r})$
and $\Vl(z)$, to yield a much smaller magnitude $\V_{\rm
  LMF}(\vect{r})$.

\begin{figure}[hbt]
  \centering 
  \includegraphics*[width=8cm]{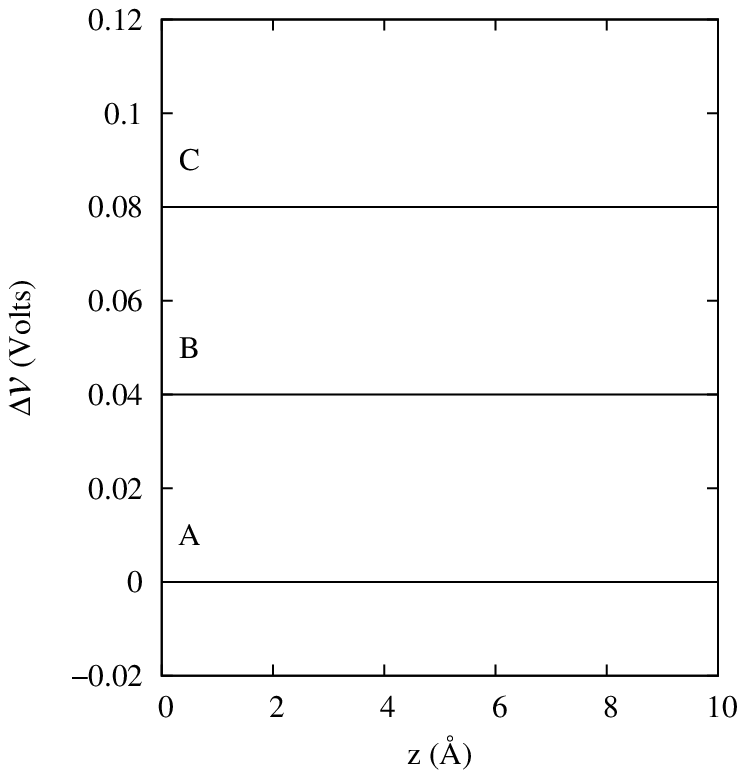}
  \caption{Plots of $\Delta \V(\vect{r})$ in Volts averaged above the sites A, B, and C as
 displayed in \fref{fig:Ptunit} for smoothing lengths $\sig=4.5$~\AA\ and $\sig=6.0$~\AA.
Data for site B is displaced vertically by 0.04 Volts and data for site C by 0.08 Volts.
In each case the value of $\Vl(z)$ is several volts, but the correction $\Delta \V(\vect{r})$ is substantially smaller.}
  \label{fig:deltaV1}
\end{figure}

In order to demonstrate this agreement more broadly, we again study
sites A, B, and C as shown in \fref{fig:Ptunit}.  \Fref{fig:deltaV1}
shows the difference
\begin{equation}
  \Delta \V(\vect{r}) \equiv \V_{\rm Ewald}(\vect{r}) - \V_{\rm LMF}(\vect{r}),
\end{equation}
laterally averaged above each hexagonal site.  This variation $\Delta
\V(\vect{r})$ is relatively small for a \sig\ of either 4.5~\AA\ or
6.0~\AA\ since either \sig\ is greater than any relevant spacings in
the surface.  Compared to the huge discrepancies that would arise from
using solely $\Vs(\vect{r})$ and neglecting the several Volt
contribution of $\Vl(z)$, these differences are miniscule.  At larger
$z$, there are slightly larger variations, which we believe are
numerical artifacts arising from our use of a sharp cutoff radius (10.5~\AA\ for $\sig
= 4.5$~\AA\ and 13.5~\AA\ for $\sig = 6.0$~\AA) in \Vs. Given the small
scale on these graphs, even very tiny errors will visibly accumulate.

We fully expect that we should be able to treat \Vr\ as
\begin{equation}
  \Vr(\vect{r}) \simeq \Vs(\vect{r}) + \Vrl(z).
\end{equation}
In fact, we expect this approximation for \Vrl\ to be even better
than for \Vl\ since the mobile charge density is included and will act
to neutralize the fixed charged sites to an extent.  The validity of
approximating the Gaussian-smoothed equilibrium charge density profile
$\rhoqs$ as solely a function of $z$ for non-electrostatic corrugated
surface confinement discussed in \sref{sxn:simplecorr} should also
hold for the mobile charge density profile in this model system with
fixed point charges.  Thus we conclude that this system should be treatable by \lmf\ 
theory with the same one-dimensional equation as used
in~\cite{RodgersWeeks.2008.Interplay-of-local-hydrogen-bonding-and-long-ranged-dipolar}
with the sole difference that $\Vs$ is a function of $\vect{r}$.

\section{Conclusions and Outlook \label{sxn:conc}}
Based on the success of \lmf\ theory in treating a variety of charged
and molecular
systems~\cite{ChenKaurWeeks.2004.Connecting-systems-with-short-and-long,ChenWeeks.2006.Local-molecular-field-theory-for-effective,RodgersKaurChen.2006.Attraction-between-like-charged-walls:-Short-ranged,DenesyukWeeks.2008.A-new-approach-for-efficient-simulation-of-Coulomb-interactions,RodgersWeeks.2008.Interplay-of-local-hydrogen-bonding-and-long-ranged-dipolar,HuRodgersWeeks..},
we believe that \lmf\ theory is a promising new approach that permits
simple minimum-image simulations of $1/r$ interactions while still
accurately accounting for the net additive long-ranged forces using
the restructured external electrostatic potential \Vr.  While the
systems treated thus far have been simple ionic and molecular systems,
the \lmf\ framework is much more broadly applicable to simulation
systems of biomolecular and experimental interest.  Current research
using the \lmf\ approach in the Weeks research group includes the
collapse of model charged polymers~\cite{DenesyukUnpub} and the
behavior of acetonitrile near silica surfaces~\cite{HuUnpub}.

Here we have sought to present the statistical mechanical foundations
of the \lmf\ treatment of charged systems.  We have highlighted the
necessary but physically reasonable approximations leading to the
\lmf\ equation, and we have also derived the simple \lmf\ equation for
the electrostatic potential valid when treating point-charge models, both
ionic and molecular.

Furthermore, we have examined the useful consequences of understanding the
smoothly-varying \lmf\ electrostatic potential \Vrl\ as resulting from a
Gaussian-smoothed charge density.  In a previous
paper~\cite{RodgersWeeks.2008.Interplay-of-local-hydrogen-bonding-and-long-ranged-dipolar},
we showed that this charge density can be physically enlightening.
Here we instead show the practical utility of this smoothing.  The
convolution of the charge density with a Gaussian of width \sig\ 
allows us to solve much simpler \lmf\ equations than might originally
be believed.  For confinement into a slab geometry, regardless of
whether the confinement has nonelectrostatic or electrostatic
corrugations, the relevant \lmf\ equation will depend only on $z$ to a
very good approximation.

In future work, we will examine the application of \lmf\ theory to
bulk site-site fluids in much more detail~\cite{HuRodgersWeeks..} and
also demonstrate very simple corrections that allow much more accurate
thermodynamics from GT models~\cite{RodgersWeeks..}. 
Another area which we hope to explore in greater detail is methods for
optimally solving the \lmf\ equation during simulation.  As alluded to
in~\cite{RodgersWeeks.2008.Interplay-of-local-hydrogen-bonding-and-long-ranged-dipolar},
the writing of the \lmf\ equation as a Poisson equation with a
Gaussian smoothed charge density suggests that connection with work on
efficient Poisson solvers would be numerically helpful, and that simple approximations
may prove surprisingly accurate~\cite{DenesyukWeeks.2008.A-new-approach-for-efficient-simulation-of-Coulomb%
-interactions}.  Furthermore,
since Poisson's equation may be written as the minimum of a
functional~\cite{Jackson.1999.Classical-Electrodynamics}, using a
modified Car-Parinello
scheme~\cite{FrenkelSmit.2002.Understanding-Molecular-Simulation:-From-Algorithms}
to \emph{slowly} evolve to the \emph{averaged, equilibrium} solution
of the \lmf\ equation seems like another fruitful path to pursue.  In
our (unoptimized) simulations of \lmf\ systems, we observed a speed-up
of at least a factor of four relative to Ewald summation.  However,
such timing questions need to be studied in much greater detail once
an optimized path to \lmf\ solution is developed.

  Regardless of such possibilities for solution of the \lmf\ equation,
  \lmf\ theory is still an inherently equilibrium theory.  Given the
  success of a scaling principle of dynamics that states that
  equilibrium, dynamic properties will be well-captured in a reference
  system that reproduced static, structural
  properties~\cite{YoungAndersen.2005.Tests-of-an-approximate-scaling-principle-for-dynamics},
  many dynamical properties for equilibrium systems could possibly be
  modeled quite well in our \lmf-mapped systems.  But such a
  conjecture needs to be tested, and, regardless, much further theoretical
  work is required to develop the analog of \lmf\ theory for
  dynamically-evolving systems.

\section*{Acknowledgments}
This work was supported by NSF grant \# CHE-0517818.  JMR
acknowledges the support of the University of Maryland Chemical
Physics fellowship.  The computations were supported in part by the
University of Maryland.

\appendix
\section*{Appendix: Relation between the GT  \vs\ and Reaction Field Truncations }

The standard short-ranged RF pair interaction $v_{0,{\rm
    RF}}(|\vect{r}_{n}-\vect{r}_{t}|)$ is proportional  to the potential
energy between a unit point charge at
$\vect{r}_{t}$ and a unit point charge at $\vect{r}_{n}$ surrounded by
a uniform neutralizing spherical charge distribution terminating at
the sharp cutoff radius $R_c$ of the RF model.  Anticipating that we
may want a more symmetric description of the charge distributions at
$\vect{r}_{t}$ and $\vect{r}_{n}$, we can also relate the RF
$v_{0,{\rm RF}}$ to the difference between the potential energy of two
unit point charges at $\vect{r}_{t}$ and $\vect{r}_{n}$ and the
potential energy of a unit uniform spherical charge distribution at
$\vect{r}_{n}$ and a unit point charge at $\vect{r}_{t}$.  But the
subtracted term is not symmetric and the smoother charged-clouds
interaction~\cite{HummerSoumpasisNeumann.1994.Computer-simulation-of-aqueous-Na-Cl-Electrolytes}
$v_{0,{\rm CC}}(r)$ can be found by subtracting the potential energy
of two unit uniform spherical charge distributions at $\vect{r}_{n}$
and $\vect{r}_{t}$ (with cutoff radius $R_c/2$) from the potential
energy of two unit point charges at $\vect{r}_{t}$ and $\vect{r}_{n}$.
The functional form $v_{0,{\rm CC}}(r)$ was previously arrived at via
a less symmetric
argument~\cite{HummerSoumpasisNeumann.1994.Computer-simulation-of-aqueous-Na-Cl-Electrolytes}.
The functional form $v_{0,{\rm CC}}(r)$ is different than that of
$v_{0,{\rm RF}}(r)$.

In contrast, the $v_{0}(|\vect{r}_{n}-\vect{r}_{t}|) =
\erfc(|\vect{r}_{n}-\vect{r}_{t}|/\sig)/|\vect{r}_{n}-\vect{r}_{t}|$
chosen for \lmf\ theory can be related to either the energy between a
unit point charge at $\vect{r}_{t}$ and a unit point charge at
$\vect{r}_{n}$ surrounded by a neutralizing Gaussian charge
distribution of width $\sigma$ or equivalently to the difference
between the potential energy of two unit point charges at
$\vect{r}_{t}$ and $\vect{r}_{n}$ and the potential energy of two unit
Gaussian charge distributions of width $\sigma/\sqrt{2}$ at
$\vect{r}_{n}$ and
$\vect{r}_{t}$~\cite{ChenKaurWeeks.2004.Connecting-systems-with-short-and-long}.
The same functional form $\vs = \erfc(r/\sig)/r$ results in either
case.


\begin{thebibliography}{10}
\bibitem{oldfootnote} Since we have not bounded errors for the
  approximations in developing \lmf\ theory, this is an approximate
  lower bound on \sig, \ie\ $\sig \gtrsim \sigmin$.
\end{thebibliography}

\end{document}